\documentclass[twocolumn]{webofc}
\usepackage[varg]{txfonts}
\usepackage[dvipsnames]{xcolor}

\newcommand{\SU}[1]{\ensuremath{\mathrm{SU}( #1 )}}

\newcommand{\SpR}[1]{\ensuremath{\mathrm{Sp}( #1,\mathbb{R} )}}

\begin{document}

\title{Understanding the Effect of Chiral NN Parametrization on Nuclear Shapes From an \textit{Ab Initio Perspective}}

\author{\firstname{Kevin S} \lastname{Becker}\inst{1}\fnsep\thanks{\email{kbeck13@lsu.edu}} \and
        \firstname{Kristina D} \lastname{Launey}\inst{1}
        \and
        \firstname{Andreas} \lastname{Ekström}\inst{2} \and
        \firstname{Grigor H} \lastname{Sargsyan}\inst{3} \and 
        \firstname{Darin C} \lastname{Mumma}\inst{1}
        \and
        \firstname{Tomáš} \lastname{Dytrych}\inst{1,4} \and
        \firstname{Daniel} \lastname{Langr}\inst{5} \and  
        \firstname{Jerry P} \lastname{Draayer}\inst{1} }

\institute{Department of Physics and Astronomy, Louisiana State University, Baton Rouge, LA 70803, USA
\and
           Department of Physics, Chalmers Institute of Technology, Gothenburg, Sweden 
\and
           Facility for Rare Isotope Beams, Michigan State University, East Lansing, MI 48824, USA
\and           
           Nuclear Physics Institute of the Czech Academy of Sciences, 250 68 \v{R}e\v{z}, Czech Republic
\and
           Department of Computer Systems, Faculty of Information Technology, Czech Technical University in Prague, Prague 16000, Czech Republic
          }

\abstract{
  The \textit{ab initio} symmetry-adapted no-core shell model naturally describes nuclear deformation and collectivity, and is therefore well-suited to studying the dynamics and coexistence of shapes in atomic nuclei. For the first time, we analyze how these features in low-lying states of $^6$Li and $^{12}$C are impacted by the underlying realistic nucleon-nucleon interaction. We find that the interaction parametrization has a notable but limited effect on collective shapes in the lowest $^6$Li and $^{12}$C states, while collective structures in the excited $2^+$ state of $^{12}$C are significantly more sensitive to the interaction parameters and exhibits emergent shape coexistence.
}

\maketitle

\section{Introduction}
\label{intro}
Recent \textit{ab initio} studies of nuclear structure have revealed the critical importance of collective behavior in explaining and predicting various properties of nuclei across the chart \cite{DytrychLDRWRBB20,DytrychSBDV_PRL07,Rowe_book16,Henderson_2018,Ruotsalainen19,DreyfussLTDBDB16,heller2022new}. Despite its apparently universal significance, nuclear collectivity is not well understood from the underlying theory of elementary particle physics, including which parts of the nucleon-nucleon (NN) interaction are responsible for its emergence. The symmetry-adapted no-core shell model (SA-NCSM) is the ideal framework for probing such questions, as it naturally describes nuclear shapes and deformation within a microscopic and fully \textit{ab initio} framework \cite{LauneyDD16,LauneyMD_ARNPS21}. For the first time, we study how nuclear shapes in low-lying states are informed by the underlying interaction from an \textit{ab initio} perspective. Specifically, we examine how shape coexistence \cite{HeydeW11,PEGarrett1,PhysRevC.103.014311,physics4030048,PhysRevC.105.034341,PhysRevLett.128.252501,PhysRevC.89.031301,PhysRevLett.125.102502} and the mixing of shapes in low-lying states of $^6$Li and $^{12}$C are affected by the low energy constants (LECs) which parametrize the short-range correlations in chiral effective nucleon-nucleon potentials \cite{BedaqueVKolck02}. 

\section{Theoretical Methods}
\label{sa-ncsm}
\textit{Ab initio} approaches build upon a "first principles" foundation, meaning that the internucleon interactions are based on the properties of two or three nucleons and that they obey the symmetries and symmetry-breaking patterns of the underlying theory of quantum chromodynamics. We utilize the SA-NCSM \cite{LauneyDD16,LauneyMD_ARNPS21}, based on the NCSM concept \cite{NavratilVB00,BarrettNV13}, which solves the many-body Schrödinger equation for $A$ particles, given nucleon-nucleon (and possibly three-nucleon and more) interactions \cite{BedaqueVKolck02,EpelbaumNGKMW02,EntemM03}. The SA-NCSM produces exactly the same results as the NCSM for a specific Hamiltonian and harmonic oscillator (HO) model space, which is defined by the number of nucleons and the total spin $J$ of the nucleus, the spacing of single-particle orbits $\hbar \Omega$, and the $N_{\textrm{max}}$ maximum number of HO excitations allowed above the lowest HO-energy configuration. The key breakthrough of the SA-NCSM is that the basis is reorganized to a symmetry-adapted (SA) model space that respects the deformation-preserving \SU{3} symmetry or the shape-preserving \SpR{3} symmetry which manifest in nuclei across the chart. Although the framework uses symmetry groups to construct the basis states, calculations are not limited \textit{a priori} by any symmetry and can, if the nuclear Hamiltonian demands, accommodate significant symmetry breaking.

\begin{figure*}
\centering
\includegraphics[width=\linewidth]{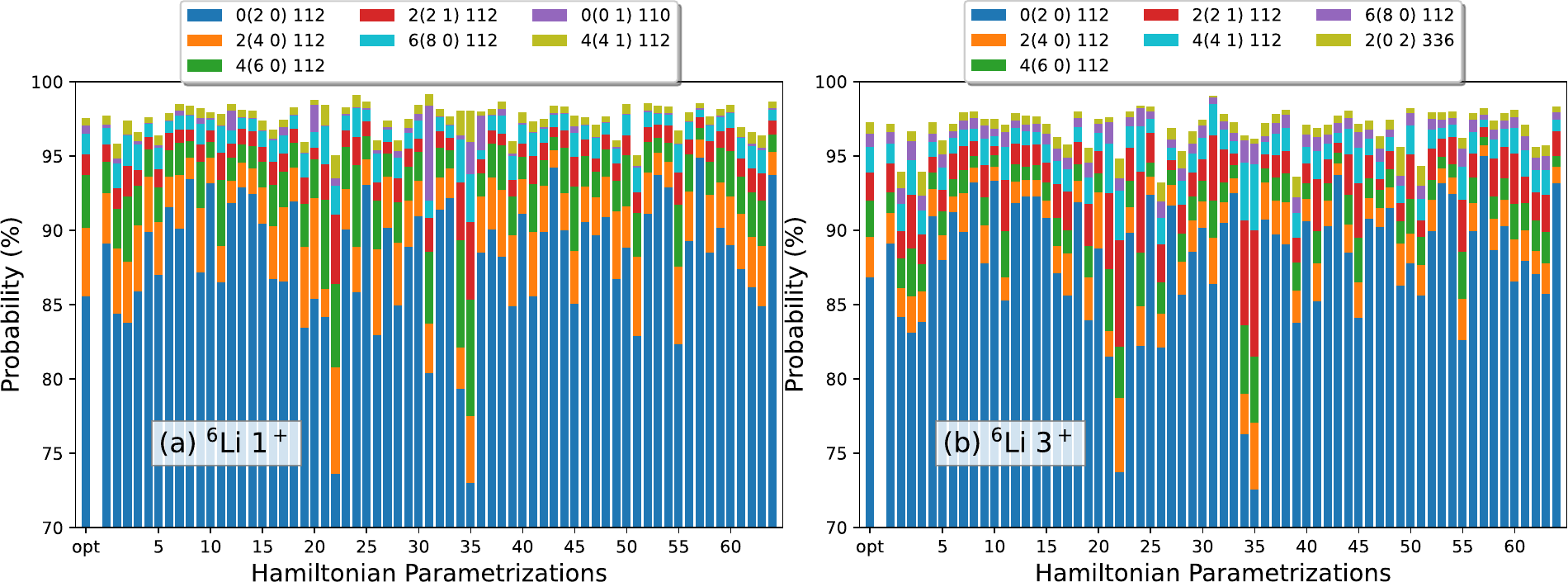}
\caption{The most dominant shapes present in (a) the $^6$Li ground state $1^+$ state in an $N_{\textrm{max}} = 10$ SA model space with 13 shapes and (b) the first excited $3^+$ state in an $N_{\textrm{max}} = 8$ SA model space with 13 shapes, obtained from 300 uniformly sampled sets of the 14 NN LECs bounded by $\pm 50\%$ their NNLO$_{\textrm{opt}}$ values (labelled as "opt"). Only the samples in which the calculated excitation energy $E_{\textrm{x}}(3^+)$ lies between $\pm 1$ MeV of the NNLO$_{\textrm{opt}}$ value are shown. The probability axes start at 70\% to better highlight the mixing of smaller configurations.}
\label{6li_wfns}
\end{figure*}

In this study, we use an \SpR{3} scheme which organizes configurations with definite \SU{3} deformation $(\lambda \, \mu)_{\omega}$ into shape-preserving subspaces (irreps) labelled by a single static deformation $N(\lambda \, \mu)_{\sigma}$, according to the group reduction chain $\SpR{3} \supset \SU{3}$ \cite{LauneyDD16,LauneyDSBD20}. Each \SpR{3}-preserving subspace is spanned by several basis states, each carrying the same $N(\lambda \, \mu)_{\sigma}$ but with different deformation, which are understood as follows. The lowest HO-energy configuration within an \SpR{3} irrep is called the bandhead or equilibrium state, and its deformation is equal to the shape quantum numbers, $(\lambda \, \mu)_{\omega} = (\lambda \, \mu)_{\sigma}$. The remaining states are given by the parity-preserving particle-hole excitations of the bandhead that preserve \SpR{3} symmetry, and are interpreted as the dynamical deformation of the bandhead, namely its surface vibrations. The quantum number $N_{\sigma}$ of a given shape indicates the total number of HO quanta in the bandhead on top of the valence configuration of the nucleus in the shell model. Hence a shape $0(\lambda \, \mu)_{\sigma}$ starts at the HO energy of the valence configuration and contains states up through $N_{\rm max}$, while $2(\lambda \, \mu)_{\sigma}$ corresponds to a shape that starts with two additional HO-energy quanta. The basis states are constructed using efficient group-theoretical algorithms, and when designing an SA model space the symplectic excitations within each irrep are cutoff at some $N_{\textrm{max}}$. The quantum numbers $(\lambda \, \mu)$ map exactly onto the macroscopic deformation parameters $\beta$ and $\gamma$, with $(\lambda \, \mu)_{\sigma} = (0 \, 0)$ indicating a spherical shape, $(\lambda \, \mu)_{\sigma} = (\lambda \, 0)$ a prolate shape, and $(\lambda \, \mu)_{\sigma} = (0 \, \mu)$ an oblate shape, with many generally triaxial shapes in between (see, e.g., \cite{heller2022new}). 

State-of-the-art SA-NCSM calculations reveal a ubiquitous pattern within low-lying nuclear states \cite{DytrychLDRWRBB20,DytrychLMCDVL_PRL12,LauneyDSBD20,DreyfussLTDB13}. Namely, out of the many shapes that span a typical no-core shell-model space, a very small subset account for an overwhelming fraction of the wave function, with one or two playing a highly dominant role. The SA-NCSM thus provides a physically-motivated method of reaching otherwise ultra-large model spaces with SA model spaces of considerably reduced sizes, which retain the accuracy of the solutions. Typically, we perform a large-scale SA-NCSM calculation without any selection up to the largest possible $N_{\textrm{max}}^C$ to determine the most significant contributions to the wave function. We then construct an SA model space by including all configurations up to some smaller $N<{N}_{\textrm{max}}^C$, and systematically extend the most dominant shapes by including more particle-hole excitations up to very large $N_{\textrm{max}}$'s that would otherwise be completely inaccessible ($N < N_{\textrm{max}}^C < N_{\textrm{max}}$).

Within this approach we employ realistic nucleon-nucleon interactions obtained from chiral effective field theory (for a detailed review, refer to \cite{Machleidt_2011}). Such potentials are derived from a perturbative expansion of the effective quantum chromodynamics Lagrangian, inherently tying them to the symmetries and symmetry-breakings of elementary particle physics. The expansion is organized by powers of the ratio of the external momentum of the interacting particles to the assumed breakdown scale of the effective theory, and by truncating the expansion one arrives at the effective nuclear forces at a given order. These forces depend on fundamental constants such as the pion mass as well as unknown parameters that must be fit to few-nucleon experimental data, ideally scattering phase shifts. These parameters are typically referred to as low energy constants (LECs), and parametrize the unresolved dynamics of quarks and gluons at low energies (see, e.g., \cite{PhysRevC.91.051301,PhysRevC.102.054301}).

At leading order (LO) in the expansion, the main contribution to the one-pion exchange interaction emerges, along with two short-ranged LEC-dependent repulsive NN contact forces acting in the $S$-wave channel which for the $^1S_0$ partial wave we split into proton-proton, proton-neutron, and neutron-neutron components for a total of four LECs. At next-to-leading order (NLO), seven LEC-dependent contacts in $S$- and $P$-waves arise, in addition to leading contributions of the medium-ranged two-pion exchange attraction. At next-to-next-to-leading order (NNLO), important sub-leading corrections are added to the two-pion exchange, introducing three more LECs, for a total of fourteen up through NNLO, in the two-nucleon sector. Three-nucleon interactions also emerge at NNLO, but for the purposes of this study they are neglected.

\section{Results and Discussion}
\label{results}

\begin{figure}[h]
\centering
\includegraphics[width=\linewidth]{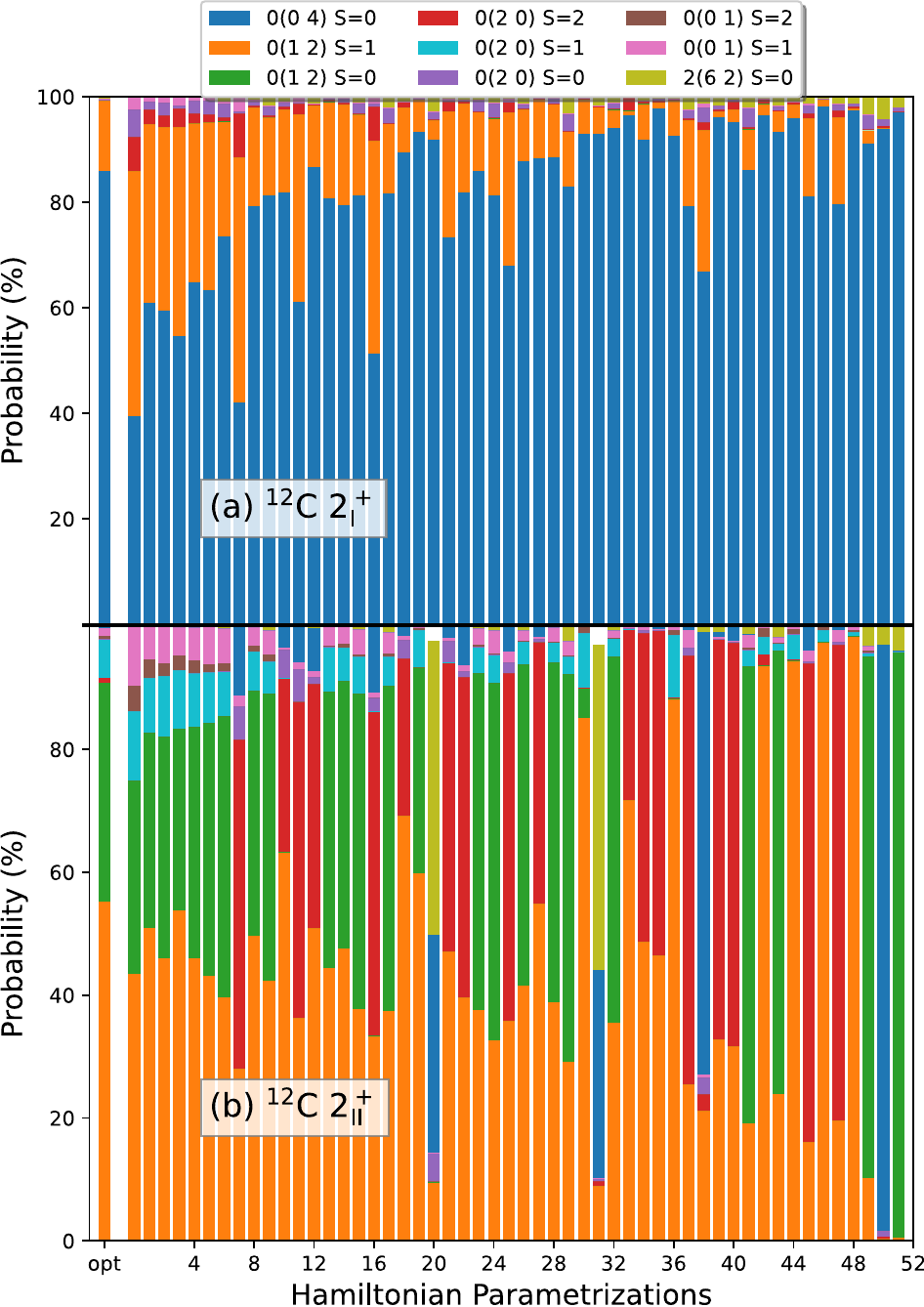}
\caption{The most dominant shapes present in the calculated (a) $^{12}$C first excited $2_{\textrm{I}}^+$ state and (b) second excited $2_\textrm{II}^+$ in an $N_{\textrm{max}} = 6$ SA model space with 14 shapes, obtained from 300 uniformly sampled sets of the NN LECs bounded by $\pm 50\%$ their NNLO$_{\textrm{opt}}$ values (labelled as "opt"). Only the samples in which the calculated energy difference $E(2_{\textrm{II}}^+) - E(2_{\textrm{I}}^+)$ lies between $\pm 5$ MeV of the NNLO$_{\textrm{opt}}$ value are shown. Both the $2_{\textrm{I}}^+$ and $2_{\textrm{II}}^+$ samples are ordered according to decreasing information entropy in the $2_{\textrm{II}}^+$ state (see FIG. 3).}
\label{12C_wfns}
\end{figure}

We study nuclear shapes by exploring, for the first time, how the symplectic content of various states and their information entropy are sensitive to the Hamiltonian parameters. Using the \SpR{3} basis, we construct SA model spaces for the $^6$Li $1^+$ ground state, its first excited $3^+$ state, and the low-lying excited $2^+$ states of $^{12}$C as follows\footnote{We note that the second $2^+_{\textrm{II}}$ state we report is labeled as second according to the SA-NCSM calculations, and does not reflect the order in the $^{12}$C experimental spectrum.}. For $^6$Li, we take 13 of the most dominant shapes according to complete $N_{\textrm{max}} = 12$ SA-NCSM calculations. In this study, for the ground state, we use 13 shapes up to $N_{\textrm{max}} = 10$. For the $3^+$, we use 13 shapes up to $N_{\textrm{max}} = 8$, and additionally include all configurations allowed at $N = 0$ and $N = 2$. For $^{12}$C, we take all 12 shapes that start at $N = 0$, as well as the $2(6 \, 2)$ and $4(12 \, 0)$ \SpR{3} irreps needed to describe the Hoyle state \cite{DreyfussLTDB13}, and include all of the allowed symplectic particle-hole excitations up to $N_{\textrm{max}} = 6$. We compute the matrix elements of all of the LEC-dependent pieces of the chiral NNLO Hamiltonian separately in these three spaces, allowing us to independently vary the strength of each term (e.g., each LEC) and compute the total Hamiltonian for different combinations of the LECs. For the two $^6$Li states, we bound each of the 14 NN LECs by $\pm 50\%$ of their NNLO$_{\textrm{opt}}$ values, draw 300 uniformly distributed LEC combinations using a Latin hypercube design \cite{lhs}, construct the parametrized potentials, and obtain wave functions for the $1^+$ and $3^+$ with probability amplitudes $|w_i|^2$, with $i$ labeling each of the shapes included in the model space. For $^{12}$C, we set the three LO isospin-breaking LECs equal to their average when computing the NNLO$_{\textrm{opt}}$ wave functions, and simultaneously vary them together around $\pm 50\%$ of this value (along with the remaining LECs) to generate 300 new samples for the $2_{\textrm{I}}^+$ and $2_{\textrm{II}}^+$. We emphasize that the aim of this study is a first step towards understanding the sensitivity of the shape structures of light nuclei to comparatively large changes in the underlying forces \cite{Becker_PRL_2023}, and to probe how nuclei should respond to such variations in the underlying physics. 

We consider only the $^6$Li samples which yield an excitation energy $E_{\textrm{x}}(3^+)$ within 1 MeV of the value obtained with NNLO$_{\textrm{opt}}$, and the $^{12}$C samples that yield an energy difference $E(2_{\textrm{II}}^+) - E(2_{\textrm{I}}^+)$ within $5$ MeV of the NNLO$_{\textrm{opt}}$ value. The contributions of the seven most dominant shapes to $^6$Li vary appreciably with the LECs, but the prolate shape $0(2 \, 0)$ which dominates both the $1^+$ and $3^+$ NNLO$_{\textrm{opt}}$ wave functions contributes at least 70\% in each $1^+$ (and $3^+$) sample (FIG. 1). This speaks to the near-perfectness of the \SpR{3} symmetry in $^6$Li that emerges from the nuclear force: the results suggest that the LEC-independent part of the NN interaction (including long-ranged physics) determines which shape will dominate this nucleus, and the LEC-driven physics tunes by how much and with which other shapes this primary configuration mixes. In $^6$Li, the other shapes which mix the most are prolate with total spin = 1 (see FIG. 1, orange, green, cyan, and red bars) or, in some cases, an oblate configuration with total spin = 0 (see FIG. 1, purple bars). Our calculations reveal no competing shapes and a predominance of a single shape in the $^6$Li $1^+$ and $3^+$ states. It is also interesting to note the clear preference for prolate shapes exhibited by the system.

\begin{figure*}
\centering
\includegraphics[width=\linewidth]{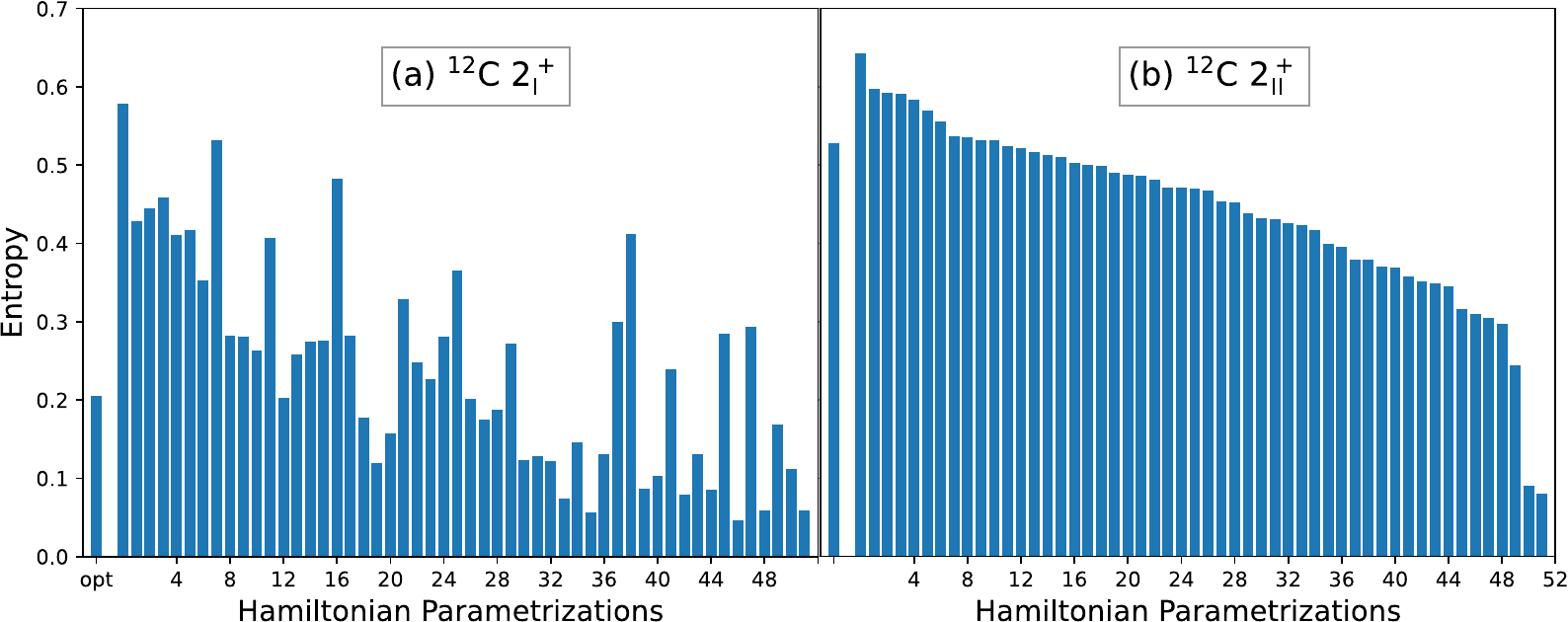}
\caption{von Neumann entropies of the (a) $^{12}$C $2_{\textrm{I}}^+$ and (b) $2_{\textrm{II}}^+$ states that yield an energy difference within 5 MeV of the NNLO$_{\textrm{opt}}$ value (labelled with "opt"), calculated from the probabilities of the 14 shapes which span the SA basis. Both sets of calculations are sorted by decreasing entropy of the second $2_{\textrm{II}}^+$ states.}
\label{entropies}
\end{figure*}

With NNLO$_{\textrm{opt}}$, the first calculated $2_\textrm{I}^+$ state of $^{12}$C is dominated primarily by one oblate $0(0 \, 4)$ shape and secondarily by a $0(1 \, 2)$ configuration with total spin = 1. We find that within these comparatively large sampling domains \cite{Becker_PRL_2023}, the $0(0 \, 4)$ almost always contributes the most to our sample wave functions, but its probability can drop as low as $40\%$ – dominant, but not overwhelmingly so as in $^6$Li (FIG. 2a, blue bars). When this probability drops, it largely mixes into the spin-1 configuration mentioned previously, shown with orange bars in FIG. 2a. The second $2_\textrm{II}^+$ state, on the other hand, displays remarkable complexity in its structure. The NNLO$_{\textrm{opt}}$ wave function consists mainly of the two spin-0 and spin-1 $0(1 \, 2)$ shapes, while the $0(0 \, 4)$ is barely present (FIG. 2b, green, orange, and blue bars). This is reflected in our sample wave functions: the $0(1 \, 2)$ configurations often contribute the most at similar probabilities, and the $0(0 \, 4)$ very minimally -- but not always. Several samples significantly enhance the contribution from a prolate $0(2 \, 0)$ with spin = 2 (see red bars in FIG. 2b), which is likely driven by a tensor force. Two samples instead simultaneously enhance the oblate $0(0 \, 4)$ and the 2-particle-2-hole (2p-2h) $2(6 \, 2)$, which is suggestive of an alpha clusterization of the system. Both of these shapes, together with the 4p-4h $4(12 \, 0)$, are known to govern the lowest three $0^+$ states in $^{12}$C, including the Hoyle state \cite{DreyfussLTDB13}, hinting that this $2_{\textrm{II}}^+$ could be associated with the third $0^+_3$ rotational band. Another interesting NN parametrization results in the first $2_{\textrm{I}}^+$ being overwhelmingly dominated by the $0(0 \, 4)$ and the second $2_{\textrm{II}}^+$ being dominated almost entirely by a $0(1 \, 2)$, indicating a highly symplectic-preserving interaction. Our results show that shape coexistence plays a driving role in the dynamics of this state, as our samples lie within a few MeV of each other with radically different shape content, often times with several shapes presenting at a comparable probability. Clearly, for the $2_{\textrm{II}}^+$ state, short-ranged correlations are responsible for determining which shape is dominant as well as the type and degree of shape coexistence.

These results can additionally be understood by examining the von Neumann information entropy $\mathcal{S}$ resulting from shape mixing in the \SpR{3}-adapted basis \cite{PhysRevC.88.044325,doi:10.1142/S0218301315300052}. The entropy is defined as $\mathcal{S}=-\sum_{i=1}^{D}\left|w_{i}\right|^{2}\log_{D}\left(\left|w_{i}\right|^{2}\right)$, where $D$ is the total number of shapes included in the model space. It is clear that no mixing ($\mathcal{S}$=0) occurs when a state contains a single shape (with $w_{j}=1$ and $w_{i\neq j}=0$) while maximal mixing ($\mathcal{S}$=1) occurs when $w_{i}=1/\sqrt{D}$ for all $i$. The entropies of the $^{12}$C samples are suppressed in the $2_{\textrm{I}}^{+}$ state, which peak at $\mathcal{S}=0.58$, compared to those in the $2_{\textrm{II}}^{+}$ state which peak at $\mathcal{S}=0.65$, since the $2_{\textrm{II}}^+$ state exhibits substantially more shape mixing than the $2_{\textrm{I}}^+$ state (FIG. 3). On the other hand, the lowest entropies occur when the oblate $0(0 \, 4)$ irrep dominates the $2_{\textrm{I}}^{+}$ state and when the $0(1 \, 2)$ shapes dominate the $2_{\textrm{II}}^{+}$, the latter of which is dramatically exemplified at sample numbers 49 and 51. 

To summarize, we find that collective features of low-lying excited $2^+$ states of $^{12}$C, such as shape coexistence and mixing, are highly sensitive to the underlying nucleon-nucleon interaction when its parameters are sampled in a relatively large window \cite{Becker_PRL_2023}, and the shape content of the nucleus can change dramatically with the chiral NNLO LECs. In contrast, we find that similar structures in low-lying $^6$Li states and the first $2^+$ state of $^{12}$C are considerably less sensitive to the LEC values, with short-ranged correlations encoded in the LEC parameters tuning the degree and type of shape mixing. 

\section{Acknowledgements}
This work was supported by the U.S. National Science Foundation (PHY-1913728, PHY-2209060), as well as in part by
the U.S. Department of Energy (DE-SC0023532,DE-SC0023694) and the Czech Science Foundation (22-14497S). This material is based upon work supported by the U.S. Department of Energy, Office of Science, Office of Nuclear Physics, under the FRIB Theory Alliance award DE-SC0013617.
This work benefited from high performance computational resources provided by LSU (www.hpc.lsu.edu), the National Energy Research Scientific Computing Center (NERSC), a U.S. Department of Energy Office of Science User Facility at Lawrence Berkeley National Laboratory operated under Contract No. DE-AC02-05CH11231, as well as the Frontera computing project at the Texas Advanced Computing Center, made possible by National Science Foundation award OAC-1818253.

\bibliography{mainfile}

\end{document}